\documentclass[%
a4paper,
prd,
twocolumn,
superscriptaddress,
preprintnumbers,
nofootinbib,
nobibnotes,
amsmath,amssymb,
aps,
floatfix
]{revtex4-2}
\usepackage{amsfonts,graphics,color}

\usepackage[english]{babel}
\usepackage{graphicx}
\usepackage{subfigure}
\usepackage{amsmath}
\usepackage{dcolumn}
\usepackage{bm}
\usepackage{comment}
\usepackage{multirow}
\usepackage{xcolor}
\definecolor{xlinkcolor}{cmyk}{1,1,0,0}
\usepackage[bookmarks=true, pdfnewwindow=true, colorlinks=true, linkcolor=xlinkcolor, citecolor=xlinkcolor, filecolor=xlinkcolor, urlcolor=xlinkcolor, final=true]{hyperref}

\usepackage[normalem]{ulem}
\definecolor{myblue}{rgb}{0.05,0.1,0.5}

\begin{document}

\preprint{INR-TH-2026-011}

\title{Bayesian Superiority in On/Off analysis}
\author{Gleb Babenov}
\email[\textbf{e-mail}: ]{babenov.gi20@physics.msu.ru}
\thanks{corresponding author}
\affiliation{Faculty of Physics, Moscow~State~University,~Leninskiye~Gory,~119991~Moscow,~Russia}
\author{Petr Satunin}
\email[\textbf{e-mail}: ]{satunin@ms2.inr.ac.ru}
\affiliation{Institute for Nuclear
Research of the Russian Academy of Sciences,\\60th~October~Anniversary~Prospect~7a,~Moscow~117312, Russia}
\affiliation{Faculty of Physics, Moscow~State~University,~Leninskiye~Gory,~119991~Moscow,~Russia}
\affiliation{Branch of Lomonosov Moscow State University in Sarov, Parkovaya~2,~607328~Sarov, Russia}

\begin{abstract}
    We present a detailed comparison of Bayesian criteria with three non-informative priors — flat, Jeffreys, and scale-invariant — for testing a signal against an unknown background and compare them with the classical frequentist Li–Ma approach in the On/Off problem. We perform Monte Carlo simulations for various background levels and evaluate the Li–Ma and Bayesian criteria by their Type I error rates. We then simulate a nonzero signal and compare the criteria in terms of Type II error rates. We find that both the Li-Ma criterion and the Bayesian criterion with the Jeffreys prior control Type I error well, excluding the case of  small background $b  <  0.5$; the Bayesian criterion with the Jeffreys prior is statistically more powerful than  the Li-Ma criterion for $b \lesssim  40$.
\end{abstract}

\maketitle

\section{Introduction}

A common problem in the statistical analysis of experimental data is the extraction of a Poisson signal from the Poisson background, which is theoretically unknown but can be measured by an independent observation in a region with {\it a-priori} no signal. This background‑measuring region is  called the Off‑source region, in contrast to the On‑source region, where the signal is looked for; the overall problem is usually called the On/Off problem.

One of the standard questions in this problem is as follows: what is the statistical significance of the signal in the On region, or with what probability can one exclude the hypothesis of no signal. The standard approach to this problem in the frequentist paradigm was performed by Li and Ma in 1983 \cite{Li:1983fv} (see also \cite{Rolke:2004mj, Cousins:2007yta}). The Li-Ma significance depends on the ratio of likelihoods of the tested hypothesis and the hypothesis with maximal likelihood estimation. The proof is based on Wilks' theorem \cite{Wilks1938}, which is asymptotic, and the method may be incorrect in the case of a small number of signal or background events.  

An alternative approach to the On/Off problem is to apply Bayesian statistics, for which a large number of events is not necessary. In the Bayesian paradigm, unknown parameters of the model are treated as random variables; the Bayesian significance is the probability for the model parameters to fit experimental data. Besides, in the Bayesian approach, a-priori information is naturally taken into account by a function of prior. Firstly, the Bayesian significance for the On/Off problem was obtained in  \cite{Gillessen:2004pm}. An analytical formula for the significance in terms of hypergeometric functions was obtained in \cite{Knoetig:2014dha, Ahnen:2015rya} for Jeffrey's prior. Priors of a wider class have been studied in  \cite{Nosek_2016,Nosek_2017}.

By performing Bayesian and frequentist methods, one obtains different values of significance, which should be compared using an objective test.  For this reason, we move from methods of calculating significance to criteria that include both the method and the rule for making a decision. The rule is similar for all methods: if the ``probability'' for a tested hypothesis is less than a given level $\alpha$, we reject it. The theory of criteria (see  e.g. the book \cite{Eadie1971}) suggests objective factors for the comparison of criteria, including, among others, Type I and Type II errors. Type I error is the probability of accepting a fluctuation of the background as a signal in the case of no true signal; Type II error refers to missing a certain signal when it actually exists. 

These definitions lead to a simple numerical test of criteria using Monte Carlo simulations.  In our work, we perform numerical simulations of the On/Off problem  at first without a signal, calculate the Type I error as a fraction of pseudo-events accepted as a signal for a given criterion, and compare it across different criteria. Further, we perform simulations with different levels of true signal and calculate Type II error as a fraction of false negative events. 
For comparison, we use the Li–Ma criterion and three Bayesian criteria with different priors: flat, Jeffreys, and scale-invariant. We find that both Li-Ma criterion and the Bayesian criterion with the Jeffreys prior control Type I error well for not small background $b > 0.5$; the Type II error for the Jeffreys criterion is statistically less than that of the Li-Ma criterion for $b < 40$.

The rest of the paper is organized as follows. In Section \ref{sec:Li-Ma} we provide the mathematical setting of the On/Off problem and show the Li-Ma solution. In Section \ref{sec:Bayes} we derive  the Bayesian criterion for three different priors. Sections \ref{sec:TypeI} and \ref{sec:TypeII} are devoted to numerically checking the criteria with Type I and Type II errors, respectively. In Section \ref{sec:over} we check the criteria for robustness against overdispersion. Sec. \ref{sec:onclusion} is devoted to the conclusion.


\section{The problem setting and frequentist Li-Ma solution}
\label{sec:Li-Ma}

We consider two independent Poisson random processes in On and Off regions, making the realizations of random variables $N_{\rm on}$ and $N_{\rm off}$ consequently; the corresponding mean values are $s+b$ and $\tau b$. Here $s$ represents signal and $b$ represents background events in the On region; $\tau$ is the ratio of the exposures of the regions, 
$$
\tau = \frac{S_{\text{off}} \, t_{\text{off}}}{S_{\text{on}} \, t_{\text{on}}}.
$$
The likelihood function is the product of two Poisson probability distributions for On and Off regions,
\begin{equation}
\label{eq:likelihood}
    L = \frac{(s + b)^{N_{\rm on}}}{N_{\rm on}!}{\rm e}^{-(s+b)}\times  \frac{(\tau b)^{N_{\rm off}}}{N_{\rm off}!}{\rm e}^{-\tau b}.
\end{equation}
Here, both signal $s$ and background $b$ are unknown parameters of the model. 
The task is to test the null hypothesis $H_0:\ s=0$ (no signal) against the alternative $H_1: s>0$ (positive signal). 

The statistic is based on the ratio of likelihood \eqref{eq:likelihood} under hypotheses $H_0$ and $H_1$, 
 \begin{equation}
 \Lambda = \frac{L(H_0)}{L(H_1)}.
 \label{eq:Lambda}
 \end{equation}
Here $L(H_0)$ is taken as eq.~\eqref{eq:likelihood} where we assumed the null hypothesis $H_0$ for signal $s=0$ while the background is determined by maximum likelihood estimation  under the condition of $H_0$,
$\hat{b}_c = \frac{1}{\tau +1}(N_{\rm on}+N_{\rm off})$.

The likelihood $L(H_1)$ for the alternative hypothesis is determined by an unconditional maximum likelihood estimation, $\hat{b} = N_{\rm off}/\tau$ and $\hat{s} = N_{\rm on} - N_{\rm off}/\tau$.

Substituting the conditional and unconditional estimations into eqs.~\eqref{eq:likelihood},~\eqref{eq:Lambda}, we obtain the expression for $-2\log \Lambda$ \cite{Li:1983fv},
  \begin{align}
    \label{eq:LiMa}
        -2\ln\Lambda &= 2\left[N_{\text{on}}\ln{\left((\tau + 1)\left(\frac{N_{\text{on}}}{N_{\text{on}} + N_{\text{off}}}\right)\right)} + \right. \notag \\ &\left.
        + N_{\text{off}}\ln{\left(\frac{\tau + 1}{\tau}\left(\frac{N_{\text{off}}}{N_{\text{on}} + N_{\text{off}}}\right)\right)}\right].
        \end{align}

The Li and Ma method is based on Wilks' theorem \cite{Wilks1938}. It states the asymptotic\footnote{It requires a large number of events, which is not always achievable in real experiments.} behavior for $-2 \ln \Lambda$, 
    \begin{equation}
    -2\ln{\Lambda} \overset{n \rightarrow \infty}{\sim} \chi_1^2,
    \label{eq:Wilks}
    \end{equation}
    where $\chi_1^2$ is the chi-squared distribution with one degree of freedom.

    Following \cite{Nosek_2016}, we introduce the  statistic:
    $$
    \text{Z}_{\text{LM}} = \text{sgn}(\hat{s})\sqrt{-2\ln{\Lambda}}.
    $$
    According to \eqref{eq:Wilks}, it follows the standard  normal distribution:
    $$
    \text{Z}_{\text{LM}} \overset{n \rightarrow \infty}{\sim} N(0, 1).
    $$
  

    To test the null hypothesis of no signal, we compare the one-sided p-value with the significance level $\alpha$. We reject $H_0$ if the p-value is less than $\alpha$, where the p-value is calculated as
    \begin{equation}
    \text{p-value} = 1 - F_{N(0, 1)}\!\left(\text{Z}_{\text{LM}}\right),
    \label{eq:p-value}
    \end{equation}
    where $F_{N(0, 1)}$ denotes the cumulative distribution function of the standard normal distribution.


\section{Bayesian criteria}
\label{sec:Bayes}



In the Bayesian paradigm, the model parameters $\lambda$ are random variables whose distribution $P(\lambda|X)$ is connected with the standard frequentist one $p(X|\lambda)$ by Bayes' formula,
    \begin{equation}
    P(\lambda|X) = \frac{p(X|\lambda)\pi(\lambda)}{\int\limits_0^{\infty}p(X|\lambda)\pi(\lambda)d\lambda},
    \label{eq:BayesFormula}
    \end{equation}
     where $X$ is experimental data,  $\pi(\lambda)$ is the prior distribution mathematically describing a-priori information.

Prior distributions can be either informative or non-informative. Here, we focus on the non-informative case; the most frequently used non-informative priors for Poisson parameters are: {\it i)} the flat (uniform) prior, $\pi(\lambda) \propto 1$; {\it  ii)} the Jeffreys prior, $\pi(\lambda) \propto 1/\sqrt{\lambda}$, obtained by Jeffreys' general rule for the Fisher information matrix \cite{Jeffreys:1939xee}; {\it iii)} the scale-invariant prior, $\pi(\lambda) \propto 1/\lambda$, which is uniform for positive $\lambda$ on a logarithmic scale, $\pi(\ln\lambda) \propto 1$, also formalized by Jeffreys \cite{Jeffreys:1939xee}. All aforementioned priors can be unified by the expression $\pi(\lambda)\propto \lambda^{-\sigma}$, where {\it  i)} corresponds to $\sigma=0$, {\it  ii)} corresponds to $\sigma=1/2$, and {\it  iii)} corresponds to $\sigma=1$.

In the On/Off problem, we take the same priors for both observations in the On and Off regions. Thus, it is more convenient to use the parameter $\lambda_{\rm on}=s+b$, which is the mean for the Poisson distribution in the On region; for the Off region, using the mean value $\tau b $ is still convenient.

 \begin{figure*}[t]
        \centering
        \includegraphics[width=0.491\linewidth]{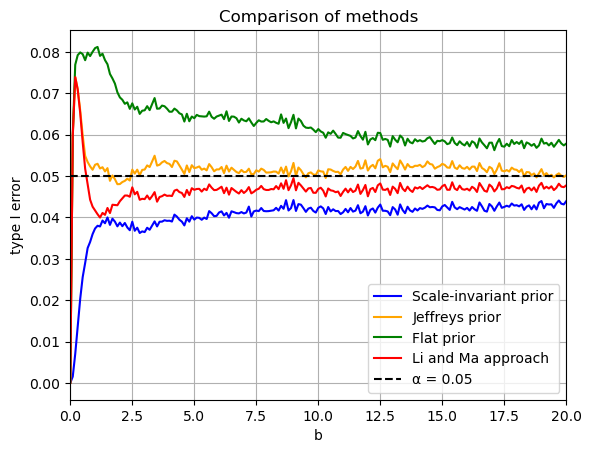}
        $\,$     \includegraphics[width=0.491\linewidth]{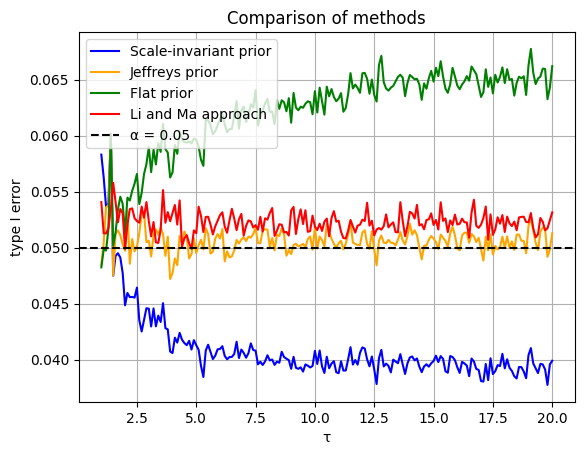}
        \caption{Type I errors for Li-Ma and three Bayesian criteria in ${\cal N} = 10^5$ Monte Carlo simulations without a signal, $s=0$. Left panel: Type I error  as a function of given background $b$ for fixed exposure ratio $\tau=5$ with a step $\Delta b=0.1$. Right panel: Type I error  as a function exposure ratio $\tau$ for fixed background $b=10$ with a step $\Delta \tau=0.1$. Significance level $\alpha=0.05$.}
        \label{fig:TypeIerror}
    \end{figure*}

Bayesian probability for non-negative random values $\lambda_{\rm on}$, $b$ with prior $\pi(\lambda)$ in the On/Off problem reads,
\begin{equation}
P(\lambda_{\rm on},b) = \frac{1}{Z} \frac{\lambda_{\rm on}^{N_{\rm on}}}{N_{\rm on}!}e^{-\lambda_{\rm on}}\frac{(\tau b)^{N_{\rm off}}}{N_{\rm off}!}e^{-\tau b}  \pi(\lambda_{\rm on})\pi(b).
\label{eq:Bayesian2Poiss}
\end{equation}
Here $Z$ is the normalization factor, see eq.~\eqref{eq:BayesFormula}. 
The condition for the presence ofa positive signal reads $s>0$, or $\lambda_{\rm on}>b$. Thus,  the corresponding Bayesian probability reads,
\begin{equation}
    P(\lambda_{\rm on}>b) = \frac{\int_0^\infty db \int_b^\infty d\lambda_{\rm on} P(\lambda_{\rm on},b)}{\int_0^\infty db \int_0^\infty d\lambda_{\rm on} P(\lambda_{\rm on},b)}.
    \label{eq:P1}
\end{equation}
Substituting Bayesian probability~\eqref{eq:Bayesian2Poiss} into the conditional expression~\eqref{eq:P1}, rescaling the integration variable $b \to b/\tau$, and performing integrations into complete and upper incomplete Gamma functions, one obtains
\begin{equation}
    P(\lambda_{\rm on}>b)\! =\! \frac{\int_0^\infty \!db\, b^{N_{\rm off}-\sigma}e^{-b}\, \Gamma(N_{\rm on} +1 - \sigma,b/\tau)}{\Gamma(N_{\rm off}+1-\sigma)\times \Gamma(N_{\rm on} +1 - \sigma)}.
    \label{eq:P3}
\end{equation}
The upper incomplete Gamma function in the numerator of eq.~\eqref{eq:P3} can be expressed as
\begin{equation}
  \Gamma(C, b/\tau) = \left(\frac{b}{\tau}\right)^C e^{-b/\tau} \int_0^\infty (1+u)^{C-1} e^{-b u / \tau} \, du,
  \label{eq:GammaInt}
\end{equation}
where $C = N_{\rm on} + 1 - \sigma$. Substituting the expression~\eqref{eq:GammaInt} into eq.~\eqref{eq:P3}, changing the order of integration in the numerator and making a change in variables $u = \frac{t\tau}{1-t} -1$, one obtains
\begin{align}
  P(\lambda_{\rm on}>b) & =  \frac{\Gamma(N_{\rm on}+N_{\rm off}+2 - 2\sigma)}{\Gamma(N_{\rm off}+1-\sigma) \Gamma(N_{\rm on} +1 - \sigma)} \times\notag \\ &\times\int_{1/(\tau+1)}^{1} t^{N_{\rm on }- \sigma} (1-t)^{N_{\rm off} - \sigma }\, dt.
\end{align}
This result can be expressed in terms of incomplete and complete beta functions,
 \begin{equation}
P(\lambda_{\rm on}>b)=
\frac{
B_{\frac{\tau}{1+\tau}}
\bigl(N_{\rm off}+1-\sigma,\; N_{\rm on}+1-\sigma \bigr)
}{
B\bigl(N_{\rm off}+1-\sigma,\; N_{\rm on} + 1-\sigma\bigr)
}.
\label{eq:P_Beta_Fun}
    \end{equation}
The expression \eqref{eq:P_Beta_Fun} for Jeffrey's prior ($\sigma=1/2$) coincides 
with the result of \cite{Knoetig:2014dha}; and for all three cases $\sigma =0,1/2,1$  with the result of \cite{Nosek_2016} where the Bayesian probability was obtained for a wider class of priors of the Gamma distribution type.

The Bayesian criterion is as follows. We reject the null hypothesis $H_0$: $s = 0$ (or $\lambda_{\rm on}>b$) at the significance level $\alpha$ if
 \begin{equation}
 P(\lambda_{\rm on}>b) >  1 - \alpha.
 \label{eq:Bayesion_crtiterion}
 \end{equation}
   This is exactly the Bayesian probability that a non-zero signal indeed exists. 


\section{Comparison of methods: Type I error}
\label{sec:TypeI}

In order to quantitatively compare Li-Ma and three Bayesian methods, we perform Monte Carlo simulations. We generate an ensemble of ${\cal N}$ On and Off events under the hypothesis $H_0$ ($s=0$)  as the realizations of the following Poisson distributions,
        $$
        N_{\text{on}} \sim \text{Pois}( b), \quad N_{\text{off}} \sim \text{Pois}(\tau b).
        $$
These events are {\it a-priori} no signal ($H_0$ is true). Nevertheless, we can apply Li-Ma \eqref{eq:LiMa},\eqref{eq:p-value} and three Bayesian \eqref{eq:P_Beta_Fun},\eqref{eq:Bayesion_crtiterion} criteria to exclude the hypothesis $H_0$ at the same significance level for the two criteria $\alpha=0.05$. The fraction of false-positive events (events for which $H_0$ is ruled out by a given method with a significance level $\alpha$) from the whole generated ensemble ${\cal N}$ is the Type I error.

\begin{figure*}[t]
             \includegraphics[width=0.3\textwidth]{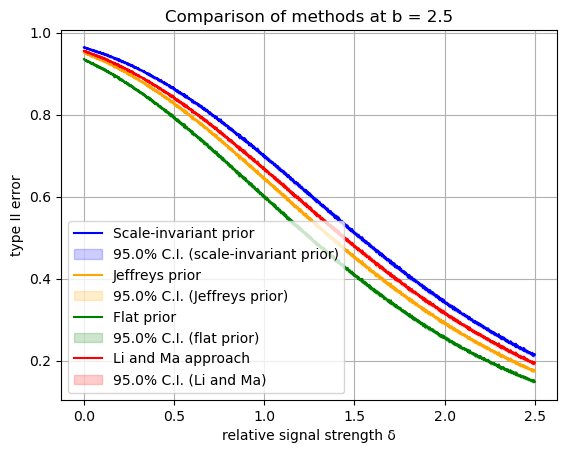}
            \includegraphics[width=0.3\textwidth]{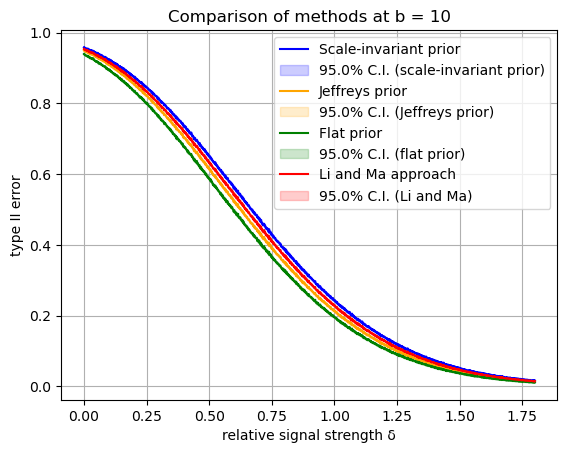}
  $\ $        \includegraphics[width=0.3\linewidth]{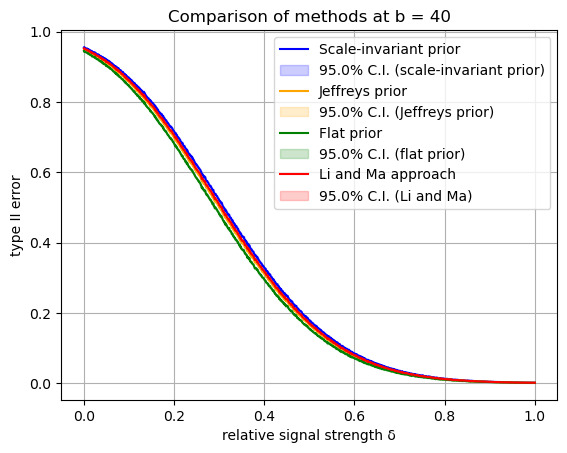}
               \includegraphics[width=0.3\textwidth]{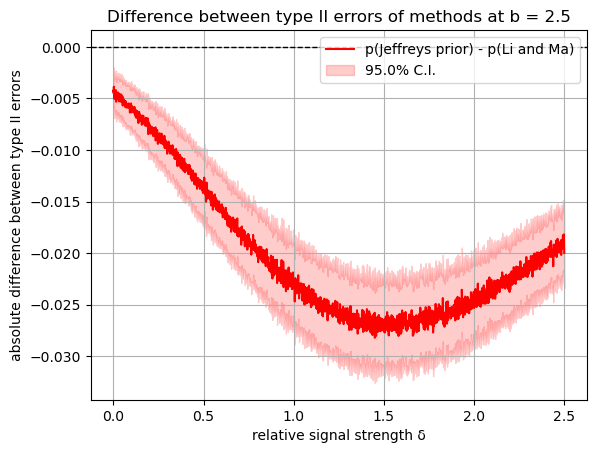}
            \includegraphics[width=0.3\textwidth]{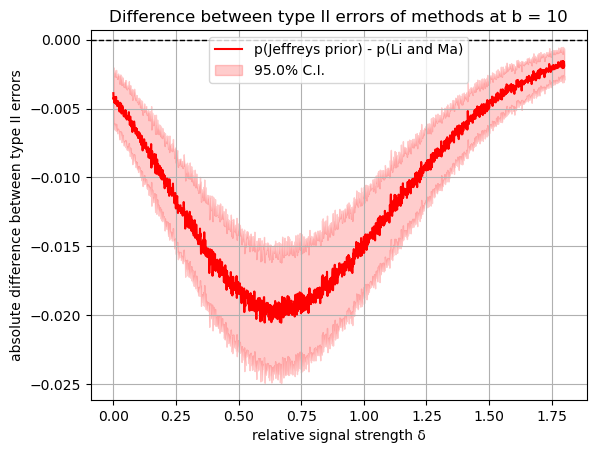}
 $\ $           \includegraphics[width=0.3\linewidth]{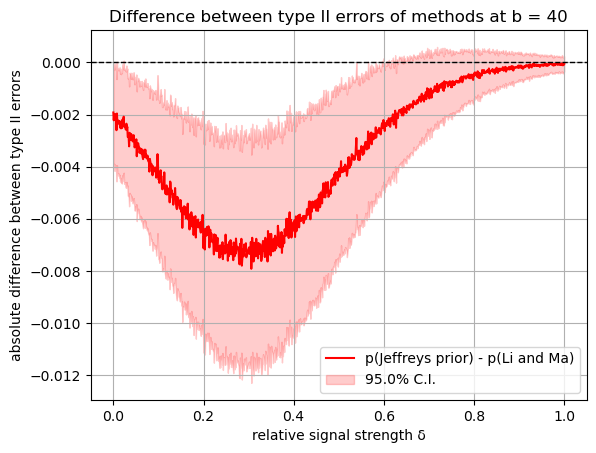}
         \caption{Type II error (upper panel) and the difference in Type II error (lower panel) for Bayesian and Li-Ma methods in ${\cal N} = 10^5$ Monte Carlo simulations with nonzero signal, as a function of relative signal strength $\delta$ with a step $\Delta \delta=10^{-3}$. $\tau=5$, $\alpha=0.05$. $b=2.5$ (left panel), $b=10$ (center panel), $b=40$ (right panel). } 
         \label{fig:TypeII}
   \end{figure*}  

The concrete simulation algorithm 
 is as follows \cite{github_repository}. We fix $\tau=5$ and vary $b$ from $b=0$ to $b=20$ with a step $\Delta b=0.1$. We generate ${\cal N} = 10^5$ events for each $b$, apply Li-Ma and Bayesian criteria with different priors for a fixed significance level $\alpha=0.05$ for each event, and draw the Type I error in Fig.~\ref{fig:TypeIerror}, left panel, for each $b$.

Fig.~\ref{fig:TypeIerror}, left panel, shows 
the following behavior of Type I error for different criteria. For large $b$, Bayesian criteria with flat and scale invariant priors tend to $\alpha=0.05$ with $b$ from different sides, but both very slowly: even for $b=20$ the over(under)estimation is still at the level of $20\%$. In contrast, Li-Ma and  Jeffrey's Type I errors are in very good agreement with  $\alpha=0.05$ for $b>2.5$. For $b < 2$, the Bayesian criterion with a flat prior shows the Type I error at a high level of $0.08$ instead of the predicted significance $\alpha = 0.05$; the similar behavior for Li-Ma and Jeffreys criteria starts for smaller $b$, $b < 0.5$. Scale invariant prior, in turn, tends to zero for small $b$. These different behaviors of the criteria are due to the significant discreteness and asymmetry of the Poisson distribution for a small mean, $b < 1$.

Fig.\ref{fig:TypeIerror}, right panel, shows the Type I error dependence on $\tau$ at fixed $b=10$.  We see that the Bayesian criteria  with flat and scale invariant priors tend to poorly estimate the significance level $\alpha$ for large $\tau$; for smaller $\tau \sim 2 - 4$, the Type I error tends to $\alpha$ with a decrease in $\tau$; however, for $\tau \simeq 1$, the fluctuations in the Type I error for all priors are quite large. 

Summarizing the aforementioned considerations, we conclude that the Li-Ma criterion and the Bayesian criterion with Jeffrey's prior control Type I error better than other criteria. The scale-invariant prior, in turn, shows less Type I error. However, as we see later, this leads to a larger Type II error.

\section{Comparison of Methods: Type II error}
\label{sec:TypeII}
        In this section, we investigate Type II error, or the probability of missing the signal when it is present (false negative result), for three Bayesian, and Li-Ma criteria at a fixed significance level $\alpha=0.05$. For this reason, we perform Monte Carlo simulations with a signal, the numerical value of which can be parameterized by relative signal strength $\delta  = s / b $.   $N_{\text{on}}$ and $  N_{\text{off}}$ are the realizations of the following Poisson distributions,
        $$
        N_{\text{on}} \sim \text{Pois}((1+\delta)\times b), \quad N_{\text{off}} \sim \text{Pois}(\tau b).
        $$
        The exposure ratio  is taken $\tau = 5$.



        We perform numerical simulations for three fixed values of background: $b=2.5$, $b=10$ and $b=40$, and for different values of relative signal strength $\delta$ \cite{github_repository}. We vary $\delta$ with a step $\Delta \delta = 10^{-3}$, and perform ${\cal N} = 10^5$ simulations for each $\delta$. Type II errors for the Bayesian and Li-Ma methods as a function of $\delta$ are presented in Fig.~\ref{fig:TypeII}, upper panel, for $b=2.5$ (left), $b=10$ (center) and $b=40$ (right). The lower panel of Fig.~\ref{fig:TypeII} shows the difference between Type II errors for Li-Ma and Bayesian methods for $b=2.5$ (left), $b=10$ (center) and $b=40$  (right), with $95\%$ CL significance.
        
        We observe that although all criteria show similar behavior, the difference is noticeable for relatively small $b$. As expected, the scale invariant (flat) prior that showed the smallest (the largest) Type I error shows the largest (the smallest) Type II error.    
        
        The difference between Li-Ma and Jeffrey's criteria is smaller but is seen with statistical $95\%$ CL, as shown in Fig.\ref{fig:TypeII}, lower panel. 
        This effect is maximal  for the intermediate signal strength:  $\delta \sim 1.5$ for $b=2.5$, $\delta \sim 0.75$ for $b=10$, and $\delta \sim 0.03$ for $b=40$, reaching the level of $0.02$ for $b=2.5$ and $b=10$ for the absolute difference of Type II errors (or statistical power) for these two methods. At larger $b$, the difference becomes smaller, as seen for $b=40$.
        As the signal increases further, both methods perform equally well, reducing the Type II error to zero at large signal strengths. 

\section{Methods Robustness for overdispersion}
\label{sec:over}

In several physical tasks, random events may not satisfy the Poisson distribution, which is characterized by variance equal to the mean, but rather a distribution with overdispersion, $\rm Var[N] > \mathbb{E}[N] $. The usual example of such a distribution is the negative binomial,
\begin{equation}
    P_{\rm NB} (N|r,p)=C^{N+r-1}_N (1-p)^N p^r,
\end{equation}
for which the mean and variance read, respectively,
\begin{equation}
   \mathbb{E}[N] = \frac{r(1-p)}{p}, \qquad \text{Var}[N] = \frac{r(1-p)}{p^2}.
\end{equation}

  \begin{figure}[t]
        \includegraphics[width=0.99\linewidth]{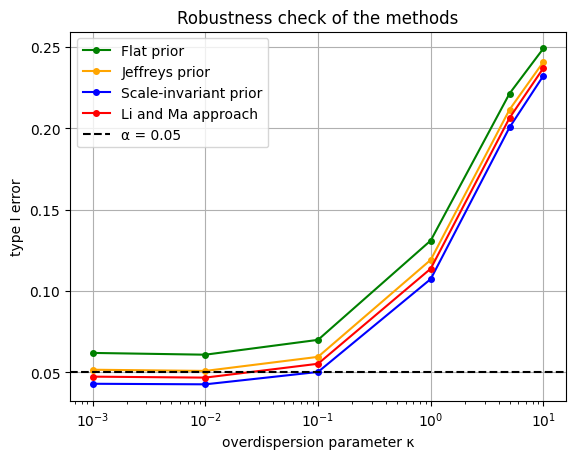}
        \caption{Dependence of Type I error on the overdispersion parameter $\kappa$ for Li-Ma and Bayesian criteria with three different priors. $b=10$, $\tau = 5$.}
        \label{overdisp}
        \end{figure}
        
In the limit $r \to \infty,\, p \to 1$ one obtains the Poisson distribution. It is more convenient to change the variables,        
$$
  r = \frac{\lambda}{\kappa}, \qquad 
        p = \frac{1}{1 + \kappa},
$$
where $\lambda$ is the mean of the distribution and $\kappa$ is the overdispersion parameter,
$$
  \mathbb{E}[N] =\lambda, \qquad \text{Var}[N] = \lambda (1 + \kappa) = \mathbb{E}[N]+\kappa \lambda.
$$
The Poisson distribution is restored in the limit $\kappa \to 0$.

 We perform a simulation of the null hypothesis $H_0$ (no signal) with a background satisfying a negative binomial distribution characterized by the overdispersion parameter  $\kappa$ ($\mathcal{N}=10^5$ events for each $\kappa$, $\kappa$ is taken on the powers of  integers from $-3$ to $1$), and we apply Li-Ma and Bayesian criteria \eqref{eq:LiMa}-\eqref{eq:p-value}, \eqref{eq:P_Beta_Fun}-\eqref{eq:Bayesion_crtiterion}, which were originally derived assuming a Poisson distribution. Type I errors for these criteria are shown in Fig.~\ref{overdisp} for $b=10,\,\tau=5$. For small $\kappa$ the Type I error coincides with Fig.~\ref{fig:TypeIerror}, left panel. The Type I error grows in comparison to the Poisson distribution starting from $\kappa = 10^{-1}$; for larger values up to $\kappa = 10$, the Type I error grows simultaneously for all criteria. 

\section{Conclusion}
\label{sec:onclusion}

Let us summarize the main idea and result of this work. First, we obtained Bayesian criteria for three different non-informative priors: uniform, scale invariant, and Jeffrey's prior, which is described by an elegant analytic formula \eqref{eq:P_Beta_Fun} that incorporates all three priors. This derivation has been fully done under the Bayesian paradigm, where the true values of signal and background are considered random variables.

Further, we have created a statistical ensemble using Monte-Carlo simulations and calculated Type I and Type II errors in strict accordance with the frequentist paradigm, as a  fraction of events in the ensemble, assuming concrete non-random values of signal $s$ and background $b$. The Bayesian approach suggests realistic formulas for the criteria, but the test is purely frequentist. 

This procedure is effective for comparing different criteria by their efficiency and power,  and it may be applied to the comparison of criteria for other tasks in statistics beyond the On/Off problem. Besides, it is intriguing if it is possible to obtain these Type I and Type II curves on analytic grounds.  

The main physical results are as follows. We have shown that  Bayesian criteria with flat and scale invariant priors overestimate and underestimate Type I error for small $b$, respectively, and  have heavy tails that very weakly tend to the significance level $\alpha$ with the growth of $b$. We have found  that  both the Li-Ma criterion and the Bayesian criterion with the Jeffreys prior control Type I error well for not small background $b > 0.5$. Moreover, the Type II error for the Jeffreys criterion is statistically less than that of the Li-Ma criterion for $b < 40$. Thus, we can recommend using the Bayesian criteria with the Jeffreys prior for On/Off analysis for relatively small beta, $b < 40$.

 



It may also be interesting to test a wider class of priors considered by \cite{Nosek_2016} as well as for informative priors.


        \paragraph{Acknowledgements} We thank Oleg Kalashev, Tatiana Kurmasheva, Frank Porter, Grigory Rubtsov and Sergey Sharakin for helpful discussions.

\bibliography{bibl}
\end{document}